%% file: Improved_Precision_in_Estimating_Average_Treatment_Effects.tex
\documentclass[12pt]{article}
\include{preamble}
\title{Improved Precision in Estimating Average Treatment Effects}
\author{Emil Pitkin, Richard Berk, Larry Brown, \\Andreas Buja, Ed George, Kai Zhang, Linda Zhao}
\begin{document}
\maketitle
\begin{abstract}
The Average Treatment Effect (ATE) is a global measure of the effectiveness of an experimental treatment intervention.  Classical methods of its estimation either ignore relevant covariates or do not fully exploit them.  Moreover, past work has considered covariates as fixed.  We present a method for improving the precision of the ATE estimate: the treatment and control responses are estimated via a regression, and information is pooled between the groups to produce an asymptotically unbiased estimate; we subsequently justify the random X paradigm underlying the result. Standard errors are derived, and the estimator's performance is compared to the traditional estimator.  Conditions under which the regression-based estimator is preferable are detailed, and a demonstration on real data is presented.
%extensions to multiple controls are included. 
\end{abstract}

\section{Introduction}
In the study of randomized controlled trials (RCTs), the average treatment effect (ATE) is a measure of an experimental intervention's global effect on a study population.  For a treatment population $T$ and control population $C$, the ATE is defined as $\tau = \expe{T} - \expe{C}$ for some measured response that can be continuous or categorical.  $\tau$ can be estimated in a multitude of ways, each estimator depending on the sampling framework and model specification. Such disparate estimators' definitions have practical significance for the researcher, who must understand the population for which his analysis holds -- that is, he must understand the scope of inference.  What's more, depending on the particular situation, the interpretation of and inference for the ATE parameter will be different.  

Past work has followed two principal strands.  The first, earliest investigations of randomized experiments centered around finite, fixed populations, all of whose members would be randomized into either treatment(s) (the number of treatments could exceed one) or control groups; the random assignment furnished the randomness, and inference extended only as far as to these subjects in the trial.  The foundation was thereby laid by Neyman, and subsequently developed by Rubin, for the notion of ``potential outcomes,'' whose unbiased estimation represented the first attempts to estimate some ATE [Neyman 1923].  Neyman considered a series of plots in a field, on each of which one of several varieties of fertilizer was applied; he wished to estimate the true average yield of the aggregated plots, even though the individual plots were fertilized with only one variety.  In this, earliest exploration of the ATE, the scope of inference was the collection of plots examined in the study only.

More recent literature has aimed to improve the precision of the ATE estimates via regression; whenever signal exists, the conditional variance of the response is reduced, with attendant gains in efficiency.  Some authors \cite{Freedman 2008a} assume the framework in which a true, generating model exists, which could be correctly and completely specified via a regression equation.  The estimating regression model in practice, however, is often misspecified, and in this case covariance adjustment can lead to undesirable consequences: in an influential critique, Freedman demonstrates how regression-based ATE estimators can lead to reduced asymptotic precision, and how they can be beset by small-sample bias.  In this paper we will step aside from the Neyman framework within which Freedman offers his analysis.

The conventional philosophy behind regression adjustments in RCTs is appealing: not only does the ATE become a parameter of the model, but the random discrepancies in empirical covariate distributions between the treatment and control groups are adjusted away, and the essential difference between treatment and control groups is retained.  In this regression analysis based RCT framework, several sources of randomness may exist. When there is a larger, target population of interest, then variation is driven by the choice of sampling units.  When inference is restricted to the sample at hand, which may not be generalizable, then randomness stems from the units' randomization to treated or control status.  Playing to the same tune as classical statistics, however, such regression analysis still presumes a fixed-X design.  Elsewhere, also in the name of improving precision of the ATE estimate, knowledge of the population mean of the covariate distribution is assumed \cite{Lin}.  

We argue for an analysis of RCTs that places minimal assumptions on the population from which data are generated. Fixed X is rarely reasonable in the context of RCTs: after patients have entered a clinical trial, nobody seriously presumes that other, putative patients in the target population have the same individual characteristics as the study subjects.  When random X is hinted at, the population mean of the covariate distribution is rarely known.  For these reasons a random-X analysis, which more realistically models experimental trials, and upon which minimal assumptions about the covariate distributions are imposed, offers the most convincing analysis.  Such an approach, with minimal assumptions placed on the data generating mechanism, echoes the work of \cite{Yang and Tsiatis} and \cite{Tsiatis and Davidian}. We will assume trials with random design, no knowledge of the covariate distributions (besides moment conditions), and will derive an efficient ATE estimator.

\section{Neyman framework}

Most pithily, the heart of Neyman's paradigm can be described as a ``repeated-sampling randomization-based'' method \cite{Rubin review}. Of N subjects $\{Y_i\}_{1:N}$, fixed once and for all, $n_T$ are assigned to the treatment group, and the remaining $n_C = N - n_T$ are exposed to the control condition. In subsequent hypothetical realizations of the experiment, another $n_T$ subjects out of the original $N$ are exposed to the treatment, and the remainder to the control.  Each of the $\binom{N}{n_T}$ subsets has an equal probability of being the ``treated block'' in any given experiment.  Note that in the thought experiment, the same, fixed $n_T$ number of units are assigned treatment, rather than each of the $n$ subjects being assigned treatment as  a Bernoulli trial with probability $\frac{n_T}{n}$.

To each subject are associated two hypothetical states, one of which is observed in practice\footnote{Of course, with multiple treatments, multiple states will be associated with each subject}. These are called ``potential outcomes,'' and they refer to the (deterministic) response of the subject, had he been subjected to the treatment (or control) condition. Let $Y_i(0)$ be the ith patient's response under the control, and let $Y_i(1)$ be the corresponding response under treatment.  The ith patient's unobserved treatment effect is defined as $Y_i(1) - Y_i(0)$. The sample-ATE, known as SATE, is defined as 

\begin{equation}
\tau^S = \frac{1}{N}\sum_{i = 1}^N [Y_i(1) - Y_i(0)] \label{SATE definition}
\end{equation}
and is estimated (WLOG let $Y_1, \ldots Y_{n_T}$ be treated) by
\begin{equation}
\hat{\tau}^S = \frac{1}{n_T}\sum_{i = 1}^{n_T} Y_i(1) - \frac{1}{n_C}\sum_{i = 1}^{n_C}Y_i(0) \label{SATE estimate}
\end{equation}
$\hat{\tau}^S$ is an unbiased estimate of $\tau^S$.  To emphasize the point, the \emph{only} source of randomness in subsequent realizations of the experiment is the subset of the original $N$ patients to whom the treatment will be assigned.  Their potential outcomes are immutable, and all that has the potential to change is which of the potential outcomes are observed.

In the literature a complementary parameter exists, called the population average treatment effect (PATE). The parameter, if the potential outcomes were known, would be computed similarly to the SATE, except the summation in \eqref{SATE definition} would be taken not over the sample in question but over all subjects in the population. In RCTs, where the desired scope of inference extends beyond the sample in question, the PATE is the more logical parameter to estimate.  The estimate will be more variable: ``sample selection error,'' defined by $\Delta_S = PATE - SATE$, adds to the uncertainty of the ATE estimate \cite{Imai 2008}.

\section{Fixed X}
The attractiveness of the Neyman framework lay in its simplicity: at its core the estimator is just a difference of means. In the name of simplicity, however, potentially useful subject specific characteristics are sacrificed.  So instead, rather than working exclusively with treatment and control responses (and taking the difference in their averages, etc.), it is possible also to estimate the ATE by way of regression.  The intention behind this approach is to make more precise the estimate of the ATE parameter by adjusting for the treated and control units' covariates.  Freedman \cite{Freedman 2008a}, responding to its pervasiveness as an estimation tool, specifically considers OLS.  He calls the ATE parameter $b_{ITT}$, where ITT is the acronym for ``intention to treat.\footnote{``Intention to treat'' is described as ``the effect of assigning everybody to treatment, minus the effect of assigning them to control.''}'' $b_{ITT}$ can be estimated via regression in several ways.  In the first, most simple and slightly contrived way, one regresses the response on the treatment indicator only, and takes note of the indicator's coefficient.  This is akin to measuring the difference of treated and control means.  For testing the equality of $b_{ITT}$ to some value, usually $0$, one employs the usual t-tests \footnote{Interestingly, the usual t-tests assume the units to have been randomly sampled, but conclusions are little affected when the assumption does not hold for a difference in means.\cite{Freedman Pisani} }

One may then proceed to introduce covariates into the regression; the new coefficient of the treatment indicator, $\hat{b}_{ITT}$, is now the estimator of $b_{ITT}$. Freedman demonstrates that while augmenting the design with covariates can improve the performance of the estimator, it can worsen it as well (standard error is either increased or decreased, depending on the data).  What's worse, the nominal standard error of $\hat{b}_{ITT}$, in addition to the estimator itself, can be severely biased.  The counterintuitive result arises because, as Freedman writes: ``randomization does not justify the assumptions behind the OLS model.'' That is, the demands the Neyman paradigm places on the nature of the data are not nearly as stringent as those imposed by OLS, what with its requirements of homoscedasticity, linearity, and fixed design. 

We, however, opt for a parallel framework, one which is not hidebound by the assumptions behind OLS.  We consider regression of a sort different from the one that Freedman critiques so compellingly.  First, he focuses in his discussion on regressions without interactions -- that is, the treatment and control groups are assumed to share slope coefficients. We will relax the assumption of homogeneous effects, and allow the treatment and control group covariates to impact to different degrees the response.   Second, in the critiqued paradigm the population of subjects is finite, and their covariates are fixed too.  The only source of randomness comes from the random assignment of treatment and control conditions. We will permit the subjects themselves (more to the point, their covariates) to be drawn from a distribution. We will describe our formulation more fully in section 4.

A recent and interesting paper by \cite{Lin} reacts to Freedman's critique, and reports the conditions under which, even in the Neyman paradigm, regression adjustment can give asymptotically valid coverage.  His most trenchant point is that, by including a full set of covariate-treatment indicator interactions in the regression model, OLS adjustment cannot worsen asymptotic precision.  Another recent paper \cite{Imbens} analyzes ATEs under more flexible circumstances, no longer working under the Neyman paradigm.  The authors present their useful results ``assuming the linear regression model is correctly specified.'' We come to similar conclusions, but under the relaxed assumptions of proper specification. 

\section{Random X formulation}

In this formulation, in contrast with the preceding discussion, nearly all quantities are random.  Whereas in the Neyman framework, only the assignment of the $n_T$ treated units is random -- but not the subject pool (hence not the covariates), nor the potential responses -- now all that will remain fixed is the number of units assigned to treatment, and the number to control.\footnote{As before, the thought experiment requires, in the next realization of the experiment, for the same $n_T$ number of subjects to be assigned treatment, and the remaining $n_C$ -- control.}  Subjects are not assigned treatment with probability $\frac{n_T}{N}$.  Mathematically, subjects are sampled independently from an infinite population; which subjects are chosen will vary from sample to sample, as will the observed covariates. 

The following discussion follows from the exposition in \cite{Buja Random X}. Let the population of subjects be described by the random variables $X_1, \ldots, X_p, Y$.  Their joint distribution $\bf{P} = \bf{P}$ $(dx_1, \ldots, dx_p, dy)$ has a full rank covariance matrix and four moments. $\Xvec$ $=(1, X_1, \ldots, X_p)'$ is the random vector of the predictor variables.  Finally, let $\mu(\Xvec)$ be the conditional mean of Y at $\Xvec$: $\mu(\Xvec) = \expe{Y | \Xvec}$.  We relax OLS assumptions, permitting, for examples, predictor variables to be omitted, and do not require -- indeed, the operating assumption is that it is not -- we do not require the true response surface to be linear in the predictors.  Instead, we work with a conditional mean that can be decomposed into linear and non-linear components.  Indeed, the linear component is thought of as the best linear approximation to the true conditional response surface; its partial slopes are defined by $\bbeta = \parens{\expe{\X\X^T}}^{-1}\expe{\X\Y}$, where the expectation is over the joint distribution of the $\X$ and the $Y$. The difference between $\mu(\Xvec)$ and $\bbeta^T\Xvec$ is denoted by $f(\Xvec)$, which is itself a random variable.  A typical decomposition of a response would look like $Y = \bbeta^T\Xvec + f(\Xvec) + \epsilon$, where the $\epsilon$ is the difference $Y - \mu(\Xvec)$. Our operating assumption is that $f(\Xvec)$ will not be uniformly equal to zero -- that is, that non-linearity will be present in the population. 

The additional results relevant to this paper are the following:
\begin{enumerate}
	\item $\bbeta$ should be estimated in the usual least squares fashion: $\hat{\bbeta} = \parens{X^TX}^{-1}X^TY$
	\item $N^{1/2}(\hat{\bbeta} - \bbeta)$ converges to a random variable with mean 0; $\hat{\bbeta}$ is an asymptotically unbiased estimator of $\bbeta$. 
	\item In finite samples, $\hat{\bbeta}$ may be a biased estimator of $\bbeta$.
	
\end{enumerate}

With the background presented, we paint in more detail the particulars of our random X formulation of how responses might be adjusted for covariates. The subjects of both the treatment and control groups are all assumed to have been sampled at random from the same population -- that is, at the population level, the covariate distributions are the same for the two groups, and assignment of treatment is independent of covariates.  The formulation is more general than in \cite{Yang and Tsiatis}, which considers a baseline measurement of $Y$ (as well as a treatment indicator) as the sole covariates. The treatment and control responses, respectively, can be denoted in the population by 

\begin{equation}
T_i = \beta^{(0)}_T + {\Xvec_{T}'}_i\bbeta_T + f_{T}(\Xvec)_i + {\epsilon_T}_i \label{Treatment regression}
\end{equation}
and, analagously, 
\begin{equation}
C_i = \beta^{(0)}_C + {\Xvec_{C}'}_i\bbeta_C + f_{C}(\Xvec)_i + {\epsilon_C}_i \label{Control regression}
\end{equation}
The $\beta^{(0)}$ are the respective intercepts at the population level, and the $\bbeta$ are the respective vectors of population partial slopes.  ${\Xvec_{T}'}$ is a random vector of treated units' covariates. Again, because we no longer assume that the response is linear in the covariates, $\beta^{(0)}_T + \Xvec_{T}'\bbeta_T$ should be thought of as the treated group's best linear approximation, at the population level, to $\expe{T|\Xvec}$. $\beta^{(0)}_T$ and $\bbeta_T$, then, are population parameters derived from population least squares regression, and minimize the expected squared distance between the linear surface and the true response surface. $f_T(\Xvec)$ is a random variable that represents the difference between the true conditional mean of $T$ and its best linear approximation in the population. In equations:

\begin{equation}
f_{T}(\Xvec) = \expe{T|\Xvec} - (\beta^{(0)}_T + \Xvec_{T}'\bbeta_T) \label{population least squares}
\end{equation}

Certain other assumptions and comments are warranted here. 

\begin{enumerate}
	\item \emph{Errors}. We place minimal demands on the errors: they should have zero mean, and be uncorrelated, conditional on the predictors. Their distributional form is unspecified, and we do not assume normality of errors. Their variances, however, we allow to differ: denote the treated and control error variances, respectively, by $\errvarT$ and $\errvarC$. 
	\item \emph{Heterogeneity} Note, also, that in the population slopes are not assumed to be the same; we allow for heterogeneous effects. The nonlinearity random variables, too, are allowed to differ between the treatment and control groups.  
	\item \emph{Equation \eqref{population least squares}}:  Moreover, $\expe{f_T \Xvec} = \zero$. The non-linear component is orthogonal to the covariates, since it is the residual from the population least squares regression.  Finally, $f_T$ should also be well-behaved, so that its variances can be assumed to exist (for example, it should be bounded, or else defined on a compact set). The same conditions apply, of course, to the control group as well.  
\end{enumerate}

As detailed in \cite{Buja Random X}, the target of estimation -- the intercept and slopes -- should be estimated, even in the random X setting, by the classical least squares estimators, and we shall do the same.

\subsection{ATE definition through regression}
We are going to re-express the ATE parameter through regression, thereby foreshadowing the proposed estimator.  As mentioned in the introduction, and using the notation developed above, the ATE is the difference between the population average of the treated subjects and their control counterparts:

\begin{equation}
\tau = \expe{T} - \expe{C} \label{difference in  means ATE definition}
\end{equation}

Subtracting \eqref{Control regression} from \eqref{Treatment regression} and taking expectations, we see that 
\begin{equation}
\tau = \parens{\beta^{(0)}_T - \beta^{(0)}_C} + \expe{\Xvec_T}\bbeta_T - \expe{\Xvec_C}\bbeta_C \label{Population regression ATE}
\end{equation}

Note that the non-linear components $f_T(\Xvec)$ and $f_C(\Xvec)$ from \eqref{Control regression} and \eqref{Treatment regression} do not appear in the equation above.  Simply, they are both equal to zero in expectation over the joint distribution of $\Xvec$ and $Y$.\footnote{This is an interesting point, whose derivation is not central to the discussion. In brief, that $\expe{f_T(\Xvec)} = 0$ follows from $\expe{f_T \Xvec} = 0$. $\Xvec$, as defined, contains an intercept; and since the expectation of the dot product of $f_T$ with a vector of ones must be zero, then $\expe{f_T \Xvec} = 0$ is equivalent to saying that $\expe{f_T} = 0 $}  It deserves mentioning that the $\bbeta$ in preceding equations are derived from the best linear approximations to the response surface, and may differ appreciably therefrom. 

The careful reader will remark that we did not simply fully, as $\expe{\Xvec_T}$ = $\expe{\Xvec_C}$ = $\expe{\Xvec}$, since, according to our assumptions, the treated and control subjects are drawn from the same population. And, indeed, \eqref{Population regression ATE} can be written as 

\begin{equation}
\tau = \parens{\beta^{(0)}_T - \beta^{(0)}_C} + \expe{\Xvec}\parens{\bbeta_T - \bbeta_C} \label{Population regression ATE one X}
\end{equation}

We consciously write these two true statement separately.  In \eqref{Population regression ATE}, one is tempted to estimate the respective expected values separately, by the respective covariate means of treatment and control groups.  In \eqref{Population regression ATE one X}, as single estimate will do, perhaps through a mean of all observed covariates, both treated and control.  There will be a difference, in practice, and we wished to emphasize it now.

One more remark: when $\expe{\Xvec} = \zero$, then $\tau = \parens{\beta^{(0)}_T - \beta^{(0)}_C}$, and the ATE is just the difference between the respective population intercepts.  This formulation hints at how we may wish to estimate the ATE from sample regressions. 

All the while, we have represented the treatment and control regressions separately, if only to emphasize that the two functional relations of covariates to the responses need bear no relation to one another in order for an ATE to be properly defined, and, later, estimated.  A single regression formulation, with interactions, may be more familiar.   The response can be written as:

\begin{equation}
	Y_i = \beta^{(0)} +  \bbeta^{(T)}I_T + \bbeta'\Xvec_i + \bbeta^{(Int)}I_T\Xvec_i + f(\Xvec)_i + I_Tg(\Xvec)_i + \epsilon_i \label{Population regression with interactions}
\end{equation}
where $g(\Xvec)$ is the difference in the treatment and control non-linearity functions.  $I_T$ is the treatment indicator at the population level; $\bbeta^{(Int)}$ is the vector in which are collected the differences in coefficients found in the treatment and control regressions respectively.  The linear approximation being the target of estimation, we will restrict our attention to estimating the $\bbeta$. In equation \eqref{Population regression with interactions} above, $\bbeta^{(T)}$ is precisely the $ATE$ parameter when the covariate expectation is equal to 0.  

\subsection{ATE estimation}
In this section we define two ATE estimators that can be derived from a random-X regression. The first reduces to the most familiar difference in means estimator, while the second borrows information across the treated and control groups.  For both estimators, we write the regression-derived expression that is equivalent to the ATE, and then appeal to plug-in MLE estimates for the associated estimator.

\begin{enumerate}
	\item Difference in means estimator. \\ 
	Recall this fact of elementary statistics: that there is one point through which the least squares regression line must pass, and that that point the mean of the predictors and the mean of the response: $\left. \hat{y} \right|_{x = \bar{x}} = \bar{y}$.  So if we substitute $\bv{\vec{\bar{X}}}_T$ into the treatment regression, the estimated conditional response will be $\bar{T}$, an unbiased estimate of $\expe{T}$. In the same way we can find an unbiased estimate of $\expe{C}$, and, as a result, of the ATE. One must be very careful when estimating the standard error of this quantity $\bracks{\hat{\beta}^{(0)}_T +  \bv{\bar{\vec{X}}}_T\hat{\bbeta}_T} - \bracks{\hat{\beta}^{(0)}_T +  \bv{\bar{\vec{X}}}_T\hat{\bbeta}_T}$, as we do in section 4.3.
	
What we have done, in effect, by substituting the respective covariate means into the separate regressions, is estimate $\expe{\Xvec}$ separately in the treatment and the control regression, which is congruent with the decomposition in \eqref{Population regression ATE}. But the winding path leads back to response sample means -- to compute them no regressions need to have been run, no covariates measured.  The lesson here is that for our purposes, controlling for covariates loses its appeal and effectiveness if no information is shared between the treatment and the control groups.

	\item A strictly regression derived estimator. \\
		Alternatively, $\expe{\Xvec}$ can -- and in most cases should -- be estimated not separately as above, twice, but rather once, by the complete set of the pooled covariates.  It should be estimated at the mean of \emph{all} covariates, $\frac{n_T\bv{\bar{\vec{X}}}_T  + n_C\bv{\bar{\vec{X}}}_C}{N}$.  The efficiency gains will be seen in section 4.2.   This approach is more congruent with \eqref{Population regression ATE one X}, so that, substituting the single estimate into \eqref{Population regression ATE}, we find that    
		\begin{equation}
			\taureg = \parens{\hat{\beta}^{(0)}_T - \hat{\beta}^{(0)}_C} + \frac{n_T\bv{\bar{\vec{X}}}_T  + n_C\bv{\bar{\vec{X}}}_C}{N}\parens{\hat{\bbeta}_T - \hat{\bbeta}_C}
\nonumber
		\end{equation}

		The estimator is invariant to location -- a shift of the empirical covariate distribution does not change the value of $\taureg$, so for the sake of appealing interpretability, we mean center the covariates.  Note that we mean-center with respect to the common, pooled mean, so that
${\parens{\Xvec_T}_i}^* = \parens{\Xvec_T}_i - \bv{\bar{\vec{X}}}$, with ${\parens{\Xvec_T}_i}^*$ defined similarly.  We thereby estimate the ATE for a covariate distribution with expectation equal to 0. From this we learn that the ATE can be estimated simply, via 

		\begin{equation}
			\taureg = \parens{\hat{\beta}^{*(0)}_T - \hat{\beta}^{*(0)}_C} \label{difference in intercepts}
		\end{equation}

		\begin{theorem} \label{taureg asymptotically unbiased}
			$\taureg$ is an asymptotically unbiased estimate of $\tau$.
		\end{theorem}

		\begin{corollary} \label{taureg unbiased}
			$\expe{\taureg} = \tau$ when
			\begin{enumerate}
				\item 
					The population response is linear in the covariates, and all covariates have been included in the statistical model, or
				\item 
					$\expe{T | X} = \expe{C | X} + k$, and $n_T = n_C$, where $k \in \reals$.
			\end{enumerate}
		\end{corollary}

That is, if the treatment and control response functions are offset by a constant, then $\taureg$ will be unbiased exactly, so long as the treatment and control sample sizes are equal.  When they are unequal, the result continues to hold when the units are inversely reweighted.  The proofs are deferred to the appendix. 

The difference in intercepts (from a mean centered regression) enriches our understanding of the relationship between a single regression with interaction terms, and one without.  In a single regression with no interactions, the ATE can be estimated via the least squares regression coefficient of the treatment indicator, which represents the constant gap between the treatment and control response surfaces. It is the difference of intercepts (that is, at $\Xvec = \zero$), but it is also the difference in responses at any arbitrary $\Xvec$ value, the difference being constant. In a single regression with interaction, the gap between the response surfaces is allowed to vary, and depends on the location of those covariates interacting with the treatment indicator.  What then, is the estimated ATE in the regression with interactions? It, too, is the coefficient of the treatment indicator.  But how else can the treatment indicator be represented and understood? It, too, is equal to the estimated difference in intercepts. Why intercepts?  Intercepts are what are left when the regression is evaluated at 0; and since we are evaluating at the average of the (pooled) mean-centered covariates, we are evaluating at $\zero$.

When
		\begin{center}
			$I_T = \left\{
			\begin{array}{ll}
			1  &  \text{Treatment is administered} \\
			0 & \text{Control is administered}
			\end{array}
		\right.$\\
		\end{center}
	
then in equations, the predicted response, when represented by a single regression with interactions, looks like
		\begin{equation}
			\hat{Y}_i = \hat{\beta}^{(0)} +  \hat{\beta}^{(T)}I_T + \Xvec\hat{\bbeta} + \Xvec\hat{\bbeta}^{(Int)}I_T \label{Regression 					with interactions estimated}
		\end{equation}

		With the covariates mean centered, substituting in the mean of the mean-centered covariates results in 

		\begin{equation}
			\left. \hat{Y}_i \right|_{\Xvec = \bv{\bar{\vec{X^*}}}} = \hat{\beta}^{(0)} +  \hat{\beta}^{(T)}I_T  \label{Regression with interactions estimated}
		\end{equation}

for which, as described, $\hat{\beta}^{(T)}$ represents the difference in intercepts. Here, the coefficient of the treatment indicator is precisely equal to $\taureg$.

\end{enumerate}

Nowhere in the definition of the model were any assumptions made about the nature of the response variables.  While a continuous response may have been implicitly assumed, the analysis is not altered if the ${T_i, C_i}$ are assumed to be count data, or to take on values ${0, 1}$. When the response is binary, the target of estimation is still $\expe{T} - \expe{C}$, but these terms can now be rewritten as $P(T) - P(C)$, where $P(T)$ represents the proportion of treatment outcomes in the population that take on the value $1$. 

One hopes that the estimate $\hat{P}(T) - \hat{C}(V)$ should fall inside $[-1, 1]$.  If one estimates $\hat{\tau}$ by the difference in means estimator, then such a desirable outcome is assured. However, $\taureg$, since it estimates the response $Y$ not at the respective sample means of the covariates ${\Xvec_i}_T$ and ${\Xvec_i}_C$ but at the weighted average $\frac{n_T\bv{\bar{\vec{X}}}_T  + n_C\bv{\bar{\vec{X}}}_C}{N}$, $\hat{P}(T) - \hat{P}(C)$ is not guaranteed with probability one to be restricted to $[-1,1]$.  The problem arises if there is limited overlap between the observed treatment and control covariates, and the slope coefficients differ appreciably between the two groups.  The probability associated with this possibility is small.

\subsection{Relative performance of ATE estimators}
We present in this section the expression for the variances of the difference-in-means and our regression based estimator, as well as for the standard error estimates, and compare the sizes of the variances.

The most familiar expression for $Var[\naive]$, of course, is $\frac{Var[T]}{n_T} + \frac{Var[C]}{n_C}$.  For the purposes of comparison to $Var[\taureg]$, the variance can be re-expressed by conditioning on covariates, and then marginalizing over their distribution, so that

\begin{lemma} \label{naive SE lemma}
	\begin{equation}
		Var\parens{\naive} =  \bracks{\frac{\sigsq_T + Var[f_T]}{n_T} + \frac{\sigsq_C + Var[f_C]}{n_C}} + \oneover{n_T}\bracks{\bbeta_T' \Sigma_{X} \bbeta_T} + \oneover{n_C}\bracks{\bbeta_C' \Sigma_{X} \bbeta_C} \label{naive_SE_in body}\\
	\end{equation}
\end{lemma}

The proof is found in the appendix.
The standard deviation of $\naive$ should be estimated by 
\begin{eqnarray} \label{naive SE estimate}
\hat{SE}(\naive) &=& \sqrt{\frac{MSE_T}{n_T} + \frac{MSE_C}{n_C} + \oneover{n_T}\parens{\hat{\bbeta}_T\hat{\Sigma^{(T)}}_X\hat{\bbeta}_T} + \oneover{n_C}\parens{\hat{\bbeta}_C\hat{\Sigma^{(C)}}_X\hat{\bbeta}_C}}
\end{eqnarray}

In the above estimate, $MSE_T$ is the mean square error computed in the treatment regression, defined as usual by $MSE_T = \frac{\sum_{i=1}^n (T_i - \hat{T}_i)^2}{N - p - 1}$, and $\hat{\Sigma}_X$ is the empirical variance-covariance matrix of the complete collection of covariates. 

The mean squared error is a scaled estimate of all the variability in the response that is not captured by the linear approximation.  So the MSE is composed of two components: the estimate of the variability in the structural errors $\epsilon$, together with the variability of $f(\Xvec)$, the random variable measuring the non-linearity in the conditional mean.  

$\taureg$ also admits a clean variance expression:

\begin{lemma} \label{taureg SE}
	\begin{equation}
		Var(\taureg) = \bracks{\frac{\sigsq_T + Var[f_T]}{n_T} + \frac{\sigsq_C + Var[f_C]}{n_C}} + O(N^{-2}) + \oneover{N}(\bbeta_T - \bbeta_C)' \Sigma_{X}(\bbeta_T - \bbeta_C)
	\end{equation}
\end{lemma}
The proof is deferred to the appendix.\newline
The standard deviation of $\taureg$ should be estimated by

\begin{eqnarray}
SE(\hat{\tau}_{regression}) &=& \sqrt{\frac{MSE_T}{n_T} + \frac{MSE_C}{n_C} + \oneover{N}(\hat{\bbeta}_T - \hat{\bbeta}_C)' \hat{\Sigma}_X(\hat{\bbeta}_T - \hat{\bbeta}_C)}
\end{eqnarray}

The more interesting claim follows: the asymptotic variance of the regression-based estimator dominates the variance of the naive estimator.

\begin{theorem} \label{asymptotic variance comparison}
	\begin{equation}
		AVar(\naive) \geq AVar(\hat{\tau}_{regression})
	\end{equation}
\end{theorem}
[Larry -- would the following form be preferred: $\lim_{n \to \infty} Var(\naive) - Var(\hat{\tau}_{\text{regression}}) \geq 0]$
The proof is found in the appendix.
\twolines
To compare the relative asymptotic efficiencies of $\naive$ and $\taureg$, only their respective variances need be compared because $\naive$ is a trivially unbiased estimate of $\tau$, and, according to \ref{taureg asymptotically unbiased}, $\taureg$ is an asymptotically unbiased estimator of the ATE.
\newline

\cite{Tsiatis and Davidian} also show that the estimator based on the model with interactions -- they call it the $ANCOVA_2$ model -- is efficient, and compare it with a large class of augmentation estimators.  The aim here is to describe the nature of the interaction model's efficiency and demonstrate which terms contribute to it.  The inequality in \ref{asymptotic variance comparison} is not strict; and equality between the asymptotic variances can be attained, and is attained iff $\bbeta_C = -\frac{n_C}{n_T}\bbeta_T$.  When the treatment and control sample sizes are equal, for example, then equality is attained when $\bbeta_C = -\bbeta_T$.  In this case, when the treatment and control slopes are negative inverses of each other, the regression-based estimate of the ATE is maximally variable.  This makes sense: sample estimates of the difference in intercepts are just as likely to be positive as to be negative, with equal probabilities of linearly increasing magnitudes of difference.

Theorem 1 refers, however, to the true variance of the respective estimators, rather than to their estimated variances\footnote{Which means that in a given sample, $\hat{SE(\taureg)}$ may exceed $\hat{SE(\naive)}$}. The theorem could analogously have been written, and should be seen here for clarity, as 
\begin{equation}
\expe{\hat{Var}(\naive)} \geq \expe{\hat{Var}(\hat{\tau}_{regression})} \nonumber
\end{equation}

A remark on the seemingly different estimators of $\naive$. 
Every introductory statistics textbook will teach that 
\begin{equation} \label{conventional diff variance}
	Var[\bar{T} - \bar{C}] = \frac{\var{T}}{n_T} + \frac{\var{C}}{n_C}
\end{equation}
and that it is estimated unbiasedly -- for example, for the purpose of hypothesis testing -- by
\begin{equation}  \label{conventional SE diff}
	\frac{{s^2}_T}{n_T} + \frac{{s^2}_C}{n_C}
\end{equation}
  In our paper, we wrote different expressions for the variance and standard error estimates of $\naive$.  This was done for ease of comparison.  In fact, \eqref{naive SE lemma} and \eqref{conventional diff variance} are equal, as are \eqref{naive SE estimate} and \eqref{conventional SE diff}, which are unbiased estimates thereof.

\subsection{Conditional and marginal estimation} We pause to make explicit the essential difference between conditional and marginal inference in our problem, and to emphasize the role of covariates that are here random.  The variance of the difference-in-means estimator is a marginal variance: over all conceivable repetitions of the experiment, as new subjects are sampled and assigned a treatment or a control condition, irrespective of any other measured or unmeasured covariates, 

\begin{equation} 
	Var[\bar{T} - \bar{C}] = \frac{Var[T]}{n_T} + \frac{Var[C]}{n_C}  \label{Classical Variance}.
\end{equation}

It is estimated, unbiasedly, by $\frac{{s^2}_T}{n_T} + \frac{{s^2}_C}{n_C}$.

Now,as in our problem, measure covariates, and run two separate regressions, so that $\hat{T} = {\hat{\beta}_T}^{(0)} +  \Xvec_{T}\hat{\bbeta}_{T}$, and 
$\hat{C} = {\hat{\beta}_C}^{(0)} + \Xvec_{C}\hat{\bbeta}_{C}$.  From elementary regression, if we estimate the response at the mean of the predictors, then $\left. \hat{T}_i \right|_{\Xvec_T = \bv{\vec{\bar{\X_T}}}} = \bar{T}$, and $\left. \hat{C}_i \right|_{\Xvec_C = \bv{\vec{\bar{\X_C}}}} = \bar{C}$.  Apparently, in estimating the ATE, $\bar{T} - \bar{C} = \left. \hat{T}_i \right|_{\Xvec_T} - \left. \hat{C}_i \right|_{\Xvec_C}$, so the variance should depend on the the observed covariates! What, then, is the proper variance of $\bar{T} - \bar{C}$?  Is it the same as that reported in \eqref{Classical Variance}? 

It will not be equal, for the simple reason that the classical variance is considered conditional on the observed covariates. To compute, note that $\bar{T}$ is independent of $\bar{C}$, so let us for the moment consider just $Var[\bar{T}]$.  $\bar{T}$ was estimated in a regression at a specific covariate value.  For ease of exposition, recall the prediction variance from simple regression, where

\begin{equation}
	\hat{Var}[\hat{y} | X = x_p] = MSE \bracks{1 + \oneover{n_T} + \frac{\parens{x_p - \bar{x}}^2}{\sum_{i = 1}^{n_T} \parens{x_i - \bar{x}}^2}}
\end{equation}
That is to say, at the covariate mean, 
\begin{equation}
	\hat{Var}[\hat{T} | \Xvec = \vec{\bar{\bv{\X}_T}}] = MSE \bracks{1 + \oneover{n_T}} \label{conditional regression variance at the mean}
\end{equation}
which, of course, does not uniformly equal $\frac{{s^2}_T}{n_T}.$ As a matter of fact, the two estimated variances will be equal only when the $R^2$ from the regression exceeds $\frac{p + 2}{n_T + 1}$, where $p$ is the number of covariates; then the regression based estimated variance will be smaller than that of the marginal, conventional estimated variance.  The reason for this discrepancy, for how the relative variances of ostensibly the same statistic depend on the quality of the fit, is simple.  

The variance estimated in \eqref{conditional regression variance at the mean} relies on classical regression theory, where the predictors are assumed to be fixed from one realization of the data to the next.  Inference is therefore conditional on the covariates; the estimate of the variance of $\bar{T}$ in \eqref{conditional regression variance at the mean} is \emph{conditional} on being estimated at the (here, fixed) mean of the covariates.  It is saying: when the mean of the covariates is equal exactly to the mean of the covariates in this sample, what is the variability of the average response? What is unaccounted for is that that selfsame covariate mean is a random quantity, and its variability will contribute to the variability in the average response.  This naive regression based estimate \eqref{conditional regression variance at the mean}, therefore, artificially deflates the true variance of the response mean.  In our analysis we compare two marginal variances, from which an inequality follows that holds for all fits.  	

\subsection{Alternative Conditions}

\subsubsection{Distribution of $\X$ known} \label{subsub:X known}
Throughout the discussion and analysis, we have assumed that the underlying distribution of $\X$ is unknown.  The alternative may present in practice where, for example, covariates like age, weight and income, for which measurements exists in the whole population, are used in the study.  In such a case, the variability inherent in estimating $\expe{\X}$ is removed (only the regression slopes remain to be estimated), with a corresponding diminution of the standard error of the ATE. The precise degree to which the standard error diminishes can be found in the appendix.

\subsubsection{Treatment Correlated with Covariates}
In the preceding discussion, we had assumed that the assignment of treatment (the treatment indicator) was independent of the covariates, with correlation among them presenting itself only in samples.  It is conceivable and natural, however, that the decision to administer treatment should depend on the covariates: perhaps, by design and because of cost constraints in the study, the researcher wishes to offer expensive treatment to a higher proportion of those suspected to require it for a shorter duration.

Precisely, suppose that the regression is written as in \ref{Population regression with interactions}, except that $I_T = H(\Xvec)$, either deterministically or stochastically, as when $I_T \sim Bern\parens{H\parens{\Xvec}}$.  The treatment indicator is a function of the covariates so the assignment mechanism is different across different strata.  In this case, the functional form of $H\parens{\cdot}$ is known, so that $\pi_i = P\parens{I_T = 1 | \Xvec}$ does not need to be estimated. 

With the goal of estimating the $ATE$, an inverse probability weighting scheme is natural because it can reduce the bias that would result from the differing sampling regimes across strata.  Accordingly, reweight the observed response $y_i$ according to

\begin{equation}
y^{(T)*}_i = \frac{y^{(T)}_i}{\pi_i}\nonumber
\end{equation}

with $\pi_i$ defined as above for the treated units, and 

\begin{equation}
y^{(C)*}_i = \frac{y^{(C)}_i}{1 - \pi_i}\nonumber
\end{equation}

Such a reweighting has been considered by, for example, \cite{Berk}, except the functional relationship between the confounders and the treatment indicator was unknown and was consequently estimated via propensity scores. Our future work will extend to cases when this functional relationship needs to be estimated. 

One proceeds with the analysis as before, running the two separate treated and control regressions, estimating the (weighted response) at the pooled mean of the covariates, and taking the difference.  Another estimate of the ATE would be $\oneover{n_T}\sumonen{y^{(T)*}_i} - \oneover{n_C}\sumonen{y^{(C)*}_i}$, what \cite{Berk} call a weighted contrast, and is the weighted variant of the difference in means estimator considered earlier. The latter is a Horvitz-Thompson type estimator (the formal H-T estimator assumes a finite population from which one samples).  The derivations and analysis relating to the weighted scheme are beyond the scope of the current paper, and will be considered in depth in a forthcoming article.

\subsubsection{Stratification}
The results described in the preceding sections make no assumptions about the nature of the covariates, which may be discrete, continuous, or both.  An interesting special case arises when, besides the treatment indicator, the other covariates represent stratum assignment, and interactions are permitted between the treatment indicators and assignment indicators. For example, subjects may be classified by treatment/control, and highest degree of educational attainment (no high school, high school, college, etc.)  The result of this pre-stratification is a two-way ANOVA layout, with interactions. In the familiar ANOVA form, the regression model may be described by 

\begin{equation}
Y_{ijk} = \mu + s_i + \tau_j + (s\tau)_{ij} + \epsilon_{ijk}
\end{equation}
$s_i$ is the ith stratum, $i = 1, \ldots I$, $\tau_j$ is the treatment effect, $j = 0, 1$ (WLOG, let j = 1 when treatment is administered), and $(s\tau)_{ij}$ is the interaction effect. Denote the number of patients in stratum i receiving regime j by $K_{ij}$. 

The difference-in-means estimator is written simply as 

\begin{equation}
\bar{\mu} = \bar{Y}_{.1.} - \bar{Y}_{.0.}
\end{equation}
and is unbiased, since $\expe{\bar{\mu}} = \expe{Y_{.1}} - \expe{Y_{.0}}$.

Now define the local ATEs, which represent the respective within-stratum ATEs by 

\beqn
ATE_i \equiv \theta_i = \expe{Y_{i1} - Y_{i0}}. 
\eeqn

The second estimator weights the per-stratum difference-in-means by the proportion of the sample found in each stratum:

\begin{equation}
\tilde{\mu} = \sum_{i = 1}^I (\bar{Y}_{i1} - \bar{Y}_{i0})*\hat{p}_i
\end{equation}
where $\hat{p}_i$ is the sample proportion of all subjects in stratum i; it is equivalently written as $\frac{K_{i+}}{K_{++}}$.  
$\expe{\tilde{\mu}} = \sum_{i=1}^{I} p_i \theta_i = \theta$, so it is also unbiased.  The estimator is unbiased under randomized assignment and under blocking since in both instances, the proportion of treated cases in a stratum is independent of the mean, and in both cases, $\expe{\hat{p_i}} = \theta_i$.  As in \cite{Miratrix}, which gives an impressive treatment of  \emph{post}-stratification in the Neyman framework, the ATE estimate here is assumed to be well-defined -- that is, the estimator is computed conditional on the event that each stratum is populated by at least one treated and one control unit. 
This second estimator just described is precisely $\taureg$.  Our results, in particular Lemma 4.4 and Theorem 4.5 continue to hold. Under slightly modified conditions, \cite{Miratrix} and \cite{Imbens 2011} show, for example, that its variance is less than that of the difference-in-means estimator, and is higher than the variance resulting from blocking (or pre-stratification) on an order of $O\parens{N^{-2}}$.

\subsection{Illustration on real data}
We present a typical application of our regression based ATE estimator on real data. We illustrate the performance of the estimator on data furnished from a classic study discussed in \cite{Lalonde} and reanalyzed in \cite{Dehejia}.  The data in question come from the National Support Work (NSW) Demonstration.  A pool of adults with economic and social problems was randomized into two groups.  The treated group was offered job training while the control group was not.  The intent of the work in \cite{Lalonde} was to compare ATE estimates from experiments to those from observational studies.  He compared the unbiased estimate of the ATE from NSW groups to an estimate drawn by comparing the treated adults to a batch of controls collected from separate comparison groups (PSID-1 and CPS-1 in his paper).  \cite{Dehejia} apply matching techniques for this comparison; relevant for our work are the 185 treated and 260 control male subjects they analyze, and which are available from the original NSW experiment.

The following covariates were adjusted for: age, education (number of years), an indicator for black, indicator for hispanic, indicator for marital status, indicator for high school degree, and earnings in 1974.  The response measured was earnings in 1978, after the job training had concluded.

In this experimental context the difference in means is equal to $4709.4$ dollars, with a standard error equal to $443.5$. The regression based method yields a point estimate of $\hat{\tau} = 4435.2$ dollars, with an SE estimate of $431.9$. The gain in SE amounts to $3.1\%$, this when the $R^2$ of the regression of reservation price on covariates and their interaction with the treatment indicator was 0.24.  A gain of this magnitude is typical for an $R^2$ of this size. Higher $R^2$ results in higher SE gains, which is vividly demonstrated in the following section.

\subsection{Illustration on simulated data}

The datasets on RCTs we have encountered have come with an $R^2$ that doesn't far exceed $0.2$.  To more vividly illustrate the results obtained in this paper, we considered the following model.   The treated and control groups were defined, respectively, by
\begin{equation}
T = 2X_1 + 3X_2 + Z_T
\end{equation}

\begin{equation}
C = X_1 + X_2 + Z_C
\end{equation}
\noindent
where $X_1 \sim Lognormal(0,1)$, $X_2 \sim Gamma(3,4)$, and $Z_T, Z_C \iid N(0,3)$. Under these conditions, $\expe{T} - \expe{C} = 2e^{1/2} - 3/2 = 1.797$.  We simulated 10,000 times, with 250 tr.eated and 250 control units in each simulation, and recorded the $R^2$ of the combined regression, as well as the ATE and SE estimates for both the difference-in-means, and for the regression based estimator considered in this paper. The average $R^2$ in the 10,000 simulation was 0.75.  Accordingly, the average $\hat{SE}\parens{\naive} = 0.676$ (with simulation SE = $0.0011$), while the average $\hat{SE}\parens{\taureg} = 0.332$ (with simulation SE = $0.0002$). Both estimators were unbiased (up to simulation granularity), with difference-in-mean and regression-based average ATEs equal to 1.798 and 1.796, respectively. Coverage of the true ATE was equal to 0.9473 and 0.949, respectively, when using $\Phi^{-1}(0.975)$ as the multiplier.  The regression based estimate naturally leads to a more powerful test.  There was nothing particular about the model chosen; similar phenomena are observed for other choices of underlying distribution.

As a final illustration, we show the relationship between the $R^2$ from the combined model and the respective standard error estimates.  $\naive$, depending only on the response, does not depend on the quality of the regression fit. $\taureg$, however, does.  10,000 simulations were again run, except the variance of $Z_T, Z_C$ was dialed from 1 to 100, with attendant decreases in the $R^2$.  The plot of $R^2$ against $\frac{\hat{SE}\parens{\taureg}}{\hat{SE}\parens{\naive}}$ is shown.  As $R^2$ decreases, the estimated standard errors converge. For high $R^2$, the $\taureg$ enjoys a dramatically lower standard error. 

\begin{figure}[htp!]
\begin{center}
\caption{$R^2$ plotted against $\frac{\hat{SE}\parens{\taureg}}{\hat{SE}\parens{\naive}}$}
\includegraphics[width=320px, height=250px]{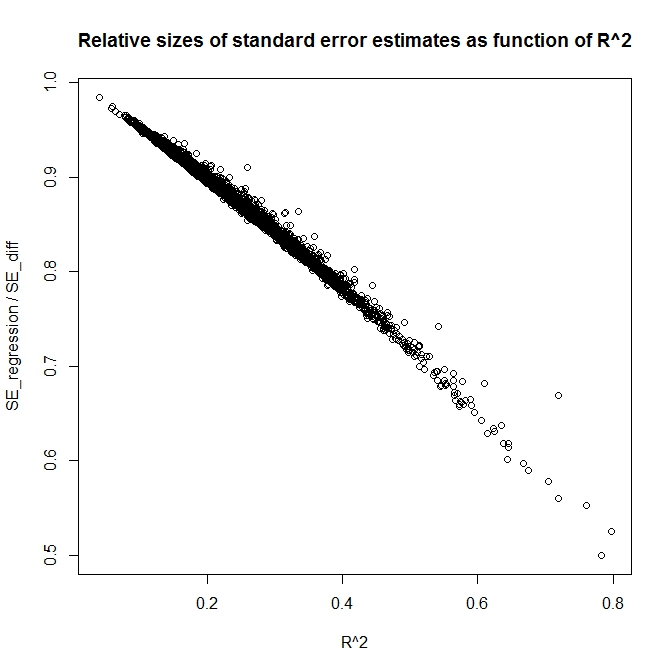}
\end{center}
\end{figure}

\section{Conclusion}
This paper lays the foundation for conducting principled and efficient asymptotic inference on ATEs. After acknowledging the aesthetics but also limitations of the Neyman paradigm, and the unreality of fixed X, we restricted our focus to an infinite population, random design, regression based estimation, where the response surface needn't be linear. Since the regression covariates are seen as random, generated from a distribution, the formulation is a more realistic representation of the practice of random sampling: randomness arises not only from the random assignment of treatment and control to subjects, but also from these subjects' (random) characteristics as well. Despite the added source of variability, the derived standard error, which takes into account these sources of randomness but also adjusts for covariates, is in expectation actually lower than its conventional counterpart. 

Bootstrapped confidence intervals can easily be generated and inference conducted for the population ATE. Moreover, the paired bootstrap, mimicking as it does the random X framework, is the natural technique for such intervals.  Future work will focus on weighting schemes when the treatment is correlated with covariates, as it would be, for example, in observational studies.  In this work we estimated with linear models.  We hope to extend the work to GLMs. 

\section{Technical appendix} 
Proof of \ref{taureg asymptotically unbiased}

After mean centering, $\taureg = \parens{\hat{\beta}^{(0)}_T - \hat{\beta}^{(0)}_C}$.  Direct application of the proposition on page 11 in \cite{Buja Random X} shows that the difference of the independent quantities $\hat{\beta}^{(0)}_T - \hat{\beta}^{(0)}_C$ is an unbiased estimate of $\beta^{(0)}_T - \beta^{(0)}_C$, which is equal to $\tau$ when $\bv{\mu} = \zero$. 

Proof of \ref{taureg unbiased}
\begin{enumerate}
	\item[(a)] When the regression model is correctly specified, then it is an introductory result that the LS estimates are unbiased: $\expe{\hat{\beta}^{(0)}_T} = \beta^{(0)}_T$ and 
	that $\expe{\hat{\beta}^{(0)}_C} = \beta^{(0)}_C$, so $\expe{\hat{\beta}^{(0)}_T - \hat{\beta}^{(0)}_C} = \beta^{(0)}_T - \beta^{(0)}_C = \tau$.
	\item[(b)] Suppose that the treatment and response surfaces have a constant offset: $n_T = n_C$ and $\expe{T|X} = \expe{C|X} + k$.  In the decomposition of $\taureg - \tau$ in the proof of \ref{taureg SE}, the only term which does not generally have expectation $\zero$ is the term denoted by $R_2$, and equal to $\bracks{\bar{\X}_T - \bar{\X}_C}\bracks{p_C(\hat{\bbeta}_T - \bbeta_T) + p_T(\hat{\bbeta}_C - \bbeta_C)}$. It will have expectation $0$ when the two bracketed terms are uncorrelated.  Exploiting the independence between the treated and control groups, the bracketed terms will be uncorrelated iff 
	\begin{equation}
		p_C Cov\parens{\bar{\X}_T, \hat{\bbeta}_T} = p_T Cov\parens{\bar{\X}_C, \hat{\bbeta}_C} \label{decorrelate the correlated}
	\end{equation}
\end{enumerate}
Inversely weight the observations, giving weight $\oneover{n_T}$ to the control observations, and $\oneover{n_C}$ to the treatment, so that \eqref{decorrelate the correlated} will hold true when $Cov\parens{\bar{\X}_T, \hat{\bbeta}_T} = Cov\parens{\bar{\X}_C, \hat{\bbeta}_C}$
When $\bbeta_C = \bbeta_T$, then, since the $\bar{\X}_T$ and $\bar{\X}_C$ are identically distributed, the above equality will hold.  $\bbeta_C = \bbeta_T$ when there is a constant offset.  

Proof of \ref{naive SE lemma}

The conventional estimator of the ATE is $\naive = \bar{T} - \bar{C}$.  Assume the covariates have zero mean; then its difference from the true ATE equals 

\begin{eqnarray}
\naive - \tau &=& \Bar{T} - \Bar{C} - \parens{\beta^{0}_T - \beta^{0}_C} \nonumber\\
&=& \bracks{\Bar{T} - \parens{\beta^{0}_T + \bar{X}_T \bbeta_T}} - \bracks{\Bar{C} - \parens{\beta^{0}_C + \bar{X}_C\bbeta_C}} \nonumber\\
&+& \bar{X}_T \bbeta_T - \bar{X}_C\bbeta_C
\end{eqnarray}

The two terms -- the former the residual means, and the latter a function of the covariates -- are independent.  Hence

\begin{eqnarray*}
Var\parens{\naive} &=& Var\braces{\bracks{\Bar{T} - \parens{\beta^{0}_T + \bar{X}_T \bbeta_T}} - \bracks{\Bar{C} - \parens{\beta^{0}_C + \bar{X}_C\bbeta_C}}}\\
&+& Var\braces{\bar{X}_T \bbeta_T - \bar{X}_C\bbeta_C}\\
&=& \bracks{\frac{\sigsq_T + Var[f_T]}{n_T} + \frac{\sigsq_C + Var[f_C]}{n_C}} + \oneover{n_T}\bracks{\bbeta_T' \Sigma_{X_T} \bbeta_T} + \oneover{n_C}\bracks{\bbeta_C' \Sigma_{X_C} \bbeta_C}\\
&=& \bracks{\frac{\sigsq_T + Var[f_T]}{n_T} + \frac{\sigsq_C + Var[f_C]}{n_C}} + \oneover{n_T}\bracks{\bbeta_T' \Sigma_{X} \bbeta_T} + \oneover{n_C}\bracks{\bbeta_C' \Sigma_{X} \bbeta_C} \label{naive_SE}\\
\end{eqnarray*}
as the covariance matrices of the treatment and control distributions are equal, since the covariates are drawn from the same distribution.  

Proof of \ref{taureg SE}

As before, we allow for unequal randomization, so that $n_T$ cases receive treatment, and $n_C$ cases receive control; denote the proportions $p_T$ and $p_C$, respectively, and suppose that $\expe{X} = \bv{\mu}$ and $Var[X] = \Sigma$. Denote the ATE by $\tau$. The ATE in the population, $\tau$, equals $\expe{T} - \expe{C} = (\beta^{0}_T - \beta^{0}_C) + \bv{\mu}(\bbeta_T - \bbeta_C)$. Then 

\beqn
\hat{\tau}_{regression} &=& \hat{\beta}^{0}_T - \hat{\beta}^{0}_C + \hat{{\bv{\mu}}}(\hat{\bbeta_T} - \hat{\bbeta_C})\\
\eeqn

\beqn
\hat{\tau}_{regression} &=& \hat{\beta}^{0}_T - \hat{\beta}^{0}_C + \bracks{p_T \bar{\X}_T + p_C \bar{\X}_C}(\hat{\bbeta_T} - \hat{\bbeta_C})\\
&=& \bar{T} - \bar{\X}_T\hat{\bbeta}_T - (\bar{C} - \bar{\X}_C\hat{\bbeta}_C) + \bracks{p_T \bar{\X}_T + p_C \bar{\X}_C}(\hat{\bbeta}_T - \hat{\bbeta}_C)\\
&=& \bar{T} - \bar{C} - \parens{\bar{\X}_T - \bar{\X}_C}\parens{p_C\hat{\bbeta}_T + p_T\hat{\bbeta}_C}
\eeqn

The multivariate mean can be taken to equal $\bf{0_p}$ WLOG since the problem is one of scale, rather than location.
So 

\begin{eqnarray}
\hat{\tau}_{regression} - \tau &=& \bar{T} - \bar{C} - \parens{\bar{\X}_T - \bar{\X}_C}\parens{p_C\hat{\bbeta_T} + p_T\hat{\bbeta_C}} - \beta^{0}_T + \beta^{0}_C \nonumber \\ 
&=& \bracks{\bar{T} - (\beta^{0}_T + \bar{\X}_T\bbeta_T)} - \bracks{\bar{C} - (\beta^{0}_C + \bar{\X}_C\bbeta_C)} \nonumber\\ 
&-& (\bar{\X}_T - \bar{\X}_C)\bracks{p_C(\hat{\bbeta}_T - \bbeta_T) + p_T(\hat{\bbeta}_C - \bbeta_C)}\nonumber\\ 
&+& (p_T\bar{\X}_T + p_C\bar{\X}_C)(\bbeta_T - \bbeta_C) \nonumber \\ 
&=& R_1 + R_2 + R_3 \label{R1 R2 R3}x
\end{eqnarray}

$R_1, R_2$, and $R_3$ are independent: $R_1$ is a function of the errors, which are independent of the covariates, while $R_2$ and $R_3$ lie in the column space of the covariates. $R_2$ is uncorrelated with $R_3$ because [we have the correlation between sums and differences of i.i.d variables. Check the math again].  Moreover, each of the terms has expectation $\bf{0_p}$ : the first, $R_1$, is a difference of average errors, equal to $\parens{\bar{\epsilon}_T + \bar{f}_T} - \parens{\bar{\epsilon}_C + \bar{f}_C}$.  The $\epsilon$ have expectation 0 by assumption, and the $f$ by construction. $R_2$ is asymptotically equal to $\zero$, for the following reason: the treatment and controls are uncorrelated, and $\expe{\bar{\X}} = \zero$, so the only component of $R_2$ not equal for all n to $\zero$ in expectation is $p_C\bar{\X}_T \hat{\bbeta}_T - p_T\bar{\X}_C \hat{\bbeta}_C$.  We'll now show that $\expe{\bar{\X}_T \hat{\bbeta}_T} \rightarrow \zero$:
\beqn
	\expe{\XbarT\betahat} &=& \expe{\XbarT\expe{\betahat | \X_T}}\\
	&=& \expe{\XbarT\parens{\X_T'\X_T}^{-1}\X_T'\expe{Y | \X_T}}\\
	&=& \expe{\XbarT\parens{\X_T'\X_T}^{-1}\X_T'\parens{\X_T\bbeta_T + f_T(\X_T)}}\\
	&=& \expe{\XbarT\parens{\X_T'\X_T}^{-1}\X_T'\X_T\bbeta_T + \XbarT\parens{\X_T'\X_T}^{-1}f_T(\X_T)}\\
	&=& \expe{\XbarT\bbeta_T} + \expe{\XbarT\parens{\X_T'\X_T}^{-1}f_T(\X_T)}\\
\eeqn
The first terms is equal to $\zero$ because $\expe{\X} = \zero$ by assumption.  The second term is equal to $\zero$ because $f_T(\X_T)$ is uncorrelated with the covariates and itself has expectation zero.

$\expe{R_3} = 0$ because $\expe{\X} = \zero$. So

\footnotesize
\begin{eqnarray}
Var(\hat{\tau}_{\text{regression}}) &=& \expe{R_1 ^2} + \expe{R_2^2} + \expe{R_3^2} \nonumber\\
&=& \braces{\parens{\expe{\bar{\epsilon}_T^2} + \expe{\bar{f}_T^2}} + \parens{\expe{\bar{\epsilon}_C^2} + \expe{\bar{f}_C^2}}} + O(N^{-2}) + (\bbeta_T - \bbeta_C)' \parens{p_T^2\frac{\Sigma_{X_T}}{n_T} + p_C^2\frac{\Sigma_{X_C}}{n_C}}(\bbeta_T - \bbeta_C) \nonumber \nonumber\\
&=& \parens{\frac{\sigsq_T}{n_T} + \frac{Var[f_T]}{n_T}} + \parens{\frac{\sigsq_C}{n_C} +\frac{Var[f_C]}{n_C}}+ O(N^{-2}) + (\bbeta_T - \bbeta_C)' \parens{p_T\frac{\Sigma_{X_T}}{N} + p_C\frac{\Sigma_{X_C}}{N}}(\bbeta_T - \bbeta_C)\nonumber\\
&=& \bracks{\frac{\sigsq_T + Var[f_T]}{n_T} + \frac{\sigsq_C + Var[f_C]}{n_C}} + O(N^{-2}) + (\bbeta_T - \bbeta_C)' \parens{\frac{\Sigma_{X}}{N}}(\bbeta_T - \bbeta_C) \label{Multi pop ATE}
\end{eqnarray}

\normalsize
The last line follows since $\Sigma_{X_T} = \Sigma_{X_C} = \Sigma_X$ -- they are all variances of a common distribution. $\blacksquare$

Proof of \ref{subsub:X known} Suppose now that the distribution of $\X$ is known.  Its mean can be assumed to be $\zero$ WLOG. Then $\tau = \beta^{0}_T - \beta^{0}_C$ and $\hat{\tau}_{regression} = \hat{\beta}^{0}_T - \hat{\beta}^{0}_C$, so that, using a similar rearrangement as before,

\begin{eqnarray}
\hat{\tau}_{regression} - \tau &=& \parens{\bar{T} - \hat{\bbeta_T}\bar{\X}_T} - \parens{\bar{C} - \hat{\bbeta_C}\bar{\X}_C} - \parens{\bbeta_T - \bbeta_C}\nonumber\\
&=& \bracks{\bar{T} - \parens{\beta^{0}_T + \bar{\X}_T\bbeta_T}} - \bracks{\bar{C} - \parens{\beta^{0}_C + \bar{\X}_C\bbeta_C}}\nonumber\\
&+& \bar{\X}_T\parens{\bbeta_T - \hat{\bbeta}_T} - \bar{\X}_C\parens{\bbeta_C - \hat{\bbeta}_C}\nonumber\\
&=& R_1 + R_2^{*} \label{R1 R2}
\end{eqnarray}

Direct comparison of \ref{R1 R2} with \ref{R1 R2 R3} will show that the estimated ATE is also asymptotically unbiased, and that its asymptotic variance is decreased by the value of $R_3$, and some of $R_2$.  With $R_3$ omitted, the standard error of the regression can just be estimated by $\sqrt{\frac{MSE_T}{n_T} + \frac{MSE_C}{n_C}}$

Proof of \ref{asymptotic variance comparison}

We now verify that the standard error of the proposed estimator dominates the standard error estimator of the conventional ATE.  We compare, therefore, 
\begin{equation}
 \bracks{\frac{\sigsq_T + Var[f_T]}{n_T} + \frac{\sigsq_C + Var[f_C]}{n_C}} + O(N^{-2}) + (\bbeta_T - \bbeta_C)' \parens{\frac{\Sigma_{X}}{N}}(\bbeta_T - \bbeta_C) \nonumber
\end{equation}

to 

\begin{equation}
\bracks{\frac{\sigsq_T + Var[f_T]}{n_T} + \frac{\sigsq_C + Var[f_C]}{n_C}} + \oneover{n_T}\bracks{\bbeta_T' \Sigma_{X} \bbeta_T} + \oneover{n_C}\bracks{\bbeta_C' \Sigma_{X} \bbeta_C}\nonumber
\end{equation}

We easily show that the asymptotic variance of the conventional estimator is higher than that of the regression estimator by comparing the variance components that differ among the two equations, noting that the $O(N^{-2})$ term vanishes. 
\begin{eqnarray}
\parens{\sqrt{\frac{n_C}{n_T}}\bbeta_T + \sqrt{\frac{n_T}{n_C}}\bbeta_C}'\Sigma_X\parens{\sqrt{\frac{n_C}{n_T}}\bbeta_T + \sqrt{\frac{n_T}{n_C}}\bbeta_C} &\geq& 0 \label{quadratic form}\\
\frac{n_C}{n_T}\parens{\bbeta_T'\Sigma_X\bbeta_T} + 2\bbeta_T'\Sigma_X\bbeta_C + \frac{n_T}{n_C}\parens{\bbeta_C'\Sigma_X\bbeta_C} &\geq& 0 \nonumber\\
\frac{N}{n_T}\bbeta_T'\Sigma_X\bbeta_T +  \frac{N}{n_C}\bbeta_C'\Sigma_X\bbeta_C &\geq&  \bbeta_T'\Sigma_X\bbeta_T - 2\bbeta_T'\Sigma_X\bbeta_C + \bbeta_C'\Sigma_X\bbeta_C \nonumber\\
\oneover{n_T}\bracks{\bbeta_T' \Sigma_{X} \bbeta_T} + \oneover{n_C}\bracks{\bbeta_C' \Sigma_{X} \bbeta_C} &\geq& (\bbeta_T - \bbeta_C)' \parens{\frac{\Sigma_{X}}{N}}(\bbeta_T - \bbeta_C) \nonumber
\blacksquare
\end{eqnarray}

The only non-algebraic step is in the first line, which is true because the LHS is a quadratic form.  Equality is attained iff $\bbeta_C = -\frac{n_C}{n_T}\bbeta_T$, which can be verified by direct substitution into \eqref{quadratic form}.

Proof of remark on $R^2$ following equation \eqref{conditional regression variance at the mean}:

$Var(\bar{T}) = \frac{SST}{n_T}$, whereas the regression based variance at the covariate mean is estimated by $MSE_T[1 + \oneover{n_T}]$, which can be rewritten as $\frac{SST-SSR}{n_T - p -1}\times\parens{\frac{n_T + 1}{n_T}}$ Dividing both expressions by $SST$ leads us to compare $\oneover{n_T}$ to $\frac{1-R^2}{n_T - p -1}\times\parens{\frac{n_T + 1}{n_T}}$.  Equality is attained when $R^2$ is equal to $\frac{p+2}{n_T+1}$

\section*{Acknowledgments}
Sincere thanks go to Paul Rosenbaum and Dylan Small for illuminating discussions, and to Dylan Small as well for recommending the Dehejia and Wahba dataset.  Thanks also to Peter Aronow for helpfully suggesting references.

%\bibliographystyle{plainnat}
%\bibliography{refs}

\end{document}

%% file: preamble.tex
\usepackage{graphicx}
\usepackage{array}
\usepackage{multirow}
\usepackage{color}
\usepackage{placeins}
\usepackage{amsmath}
\usepackage{amssymb}
\usepackage{amsfonts}
\usepackage{enumerate}
\usepackage[margin=1in]{geometry}
\definecolor{gray}{rgb}{0.5,0.5,0.5}
\definecolor{black}{rgb}{0,0,0}
\definecolor{white}{rgb}{1,1,1}
\definecolor{blue}{rgb}{0.5,0.5,1}

\definecolor{backgcode}{rgb}{0.97,0.97,0.8}
\definecolor{Brown}{cmyk}{0,0.81,1,0.60}
\definecolor{OliveGreen}{cmyk}{0.64,0,0.95,0.40}
\definecolor{CadetBlue}{cmyk}{0.62,0.57,0.23,0}

\newtheorem{theorem}{Theorem}[section]
\newtheorem{lemma}[theorem]{Lemma}

\newtheorem{corollary}[theorem]{Corollary}

\newcommand{\qed}{\nobreak \ifvmode \relax \else
      \ifdim\lastskip<1.5em \hskip-\lastskip
      \hskip1.5em plus0em minus0.5em \fi \nobreak
      \vrule height0.75em width0.5em depth0.25em\fi}

\newcommand{\bv}[1]{\boldsymbol{#1}}

\newcommand{\sigsq}{\sigma^2}

\newcommand{\betahat}{\hat{\beta}}

\newcommand{\iid}{~{\buildrel iid \over \sim}~}

\newcommand{\X}{\bv{X}}

\newcommand{\Y}{\bv{Y}}

\newcommand{\bbeta}{\bv{\beta}}

\newcommand{\reals}{\mathbb{R}}

\newcommand{\beqn}{\vspace{-0.25cm}\begin{eqnarray*}}
\newcommand{\beqnsol}{\vspace{-0.25cm}\begin{eqnarray*}\text{Solution:} \quad}
\newcommand{\eeqn}{\end{eqnarray*}}
\newcommand{\nbeqn}{\vspace{-0.25cm}\begin{eqnarray}}
\newcommand{\neeqn}{\end{eqnarray}}
\newcommand{\beqnnumb}{\vspace{-0.25cm}\begin{eqnarray}}
\newcommand{\eeqnnumb}{\end{eqnarray}}

\newcommand{\parens}[1]{\left(#1\right)}

\newcommand{\sumonen}[1]{\sum_{i=1}^n#1}

\newcommand{\bracks}[1]{\left[#1\right]}
\newcommand{\braces}[1]{\left\{#1\right\}}

\newcommand{\expe}[1]{\mathbb{E}\bracks{#1}}

\newcommand{\var}[1]{\text{Var}\bracks{#1}}

\newcommand{\oneover}[1]{\frac{1}{#1}}

\newcommand{\twolines}{\newline\newline}

\newcommand{\XbarT}{\bar{\X}_T}

\newcommand{\Xvec}{\bv{\vec{X}}}
\newcommand{\zero}{\bv{0}}
\newcommand{\naive}{\hat{\tau}_{\text{diff}}}
\newcommand{\taureg}{\hat{\tau}_{\text{regression}}}
\newcommand{\errvarT}{\sigsq_T}
\newcommand{\errvarC}{\sigsq_C}